\documentclass[showpacs,superscriptaddress,prd,floats,aps,twocolumn]{revtex4}
\usepackage{amsmath,amssymb,amsfonts}
\usepackage{graphicx}

\begin{document}
\preprint{AEI-2002-009, gr-qc/yymmddd}

\title{New conformally flat  initial data for spinning  black holes}

\author{Sergio Dain}
\affiliation{Albert-Einstein-Institut,
Max-Planck-Institut f{\"u}r Gravitationsphysik,
Am M\"uhlenberg 1, D-14476 Golm, Germany}

\author{Carlos O. Lousto}
\affiliation{Albert-Einstein-Institut,
  Max-Planck-Institut f{\"u}r Gravitationsphysik, Am M\"uhlenberg 1,
D-14476 Golm, Germany}
\affiliation{Department of Physics and Astronomy,
The  University of Texas at Brownsville, Brownsville, Texas 78520}
\affiliation{Instituto de Astronom\'{\i}a y F\'{\i}sica del Espacio--CONICET,
Buenos Aires, Argentina}

\author{Ryoji Takahashi}
\affiliation{Albert-Einstein-Institut,
Max-Planck-Institut f{\"u}r Gravitationsphysik,
Am M\"uhlenberg 1, D-14476 Golm, Germany}
\affiliation{Theoretical Astrophysics Center, Juliane Maries Vej 30,
2100 Copenhagen, Denmark}

\date{\today}

\begin{abstract}
  We obtain an explicit solution of the momentum constraint for
  conformally flat, maximal slicing, initial data which gives an
  alternative to the purely longitudinal extrinsic curvature of Bowen
  and York. The new solution is related, in a precise form, with the
  extrinsic curvature of a Kerr slice.  We study these new initial
  data representing spinning black holes by numerically solving the
  Hamiltonian constraint. They have the following features: i) Contain
  less radiation, for all allowed values of the rotation parameter,
  than the corresponding single spinning Bowen-York black hole. ii)
  The maximum rotation parameter $J/m^2$ reached by this solution is
  higher than that of the purely longitudinal solution allowing thus
  to describe holes closer to a maximally rotating Kerr one.  We
  discuss the physical interpretation of these properties and their
  relation with the weak cosmic censorship conjecture. Finally, we
  generalize the data for multiple black holes using the ``puncture''
  and isometric formulations.
\end{abstract}


\maketitle

\section{Introduction}

Black holes are expected to be common objects in the universe. At the
classical level, they are described by the Einstein field equations.
The study of black holes by an initial value formulation of
Einstein's equations is a difficult problem, both from the
analytic and numerical point of view. One of the most important open
question regarding black holes is the weak cosmic censorship
conjecture (cf. \cite{Penrose69} and also the recent review
\cite{Wald99}): generic singularities of gravitational collapse are contained
in black holes. The physical relevance of the concept of black holes
depends on the validity of this conjecture. However, a general proof
of the cosmic censorship conjecture lies outside the scope of present
mathematical techniques. Thus, it is only possible to prove it in some
very restrictive cases such spherical symmetry \cite{Christodoulou99}
or to find indirect evidence of its validity.  In the obtaining of
these indirect evidences, a key role is played by the following three
consequences of the weak cosmic censorship and the theory of black holes
\cite{Hawking73} \cite{Wald84}:

\begin{itemize}
\item[(i)] Every apparent
  horizon must be entirely contained within the black hole event horizon.
\item[(ii)] If matter satisfies
  the null energy condition (i.e. if $T_{ab}k^ak^b\geq 0$ for all null 
  $k^a$), then  the area of the event horizon of a black hole cannot
  decrease in time.
  
\item[(iii)] All black holes eventually settle down to a final Kerr
  black hole.
 
\end{itemize}

From (i)-(iii) we can deduce the Penrose inequality \cite{Penrose73} 
\begin{equation}
  \label{eq:12}
  A\leq 16\pi m^2,
\end{equation}
where $A$ is the area of the apparent horizon and $m$ is the total ADM
mass of the space-time. Remarkably, the inequality \eqref{eq:12}
involves only quantities which can be computed directly form the
initial data. After considerable effort, the Penrose inequality has
been proved for time symmetric initial data
\cite{Huisken01}\cite{Bray99}, providing an important support for the
validity of (i)-(iii). The Penrose inequality can be strengthened if
we assume axial symmetry and take into account angular momentum.
Angular momentum is a conserved quantity in axially symmetric
space-times, since it can be defined by a Komar integral (cf.
\cite{Komar59} and also \cite{Wald84}). Using this fact and (i)-(iii)
we deduce the following inequality \cite{Hawking72}
\begin{equation}
  \label{eq:13}
 \epsilon_A \leq 1, \quad \epsilon_A \equiv \frac{A}{8\pi(m^2+\sqrt{m^4-J^2})},
\end{equation}
where $J$ is the total angular momentum of the space-time. The
inequality \eqref{eq:13} must hold for every axially symmetric, non
singular, asymptotically flat initial data. The equality in
\eqref{eq:13} must be achieved if and only if the data are slices of
Kerr. Note that \eqref{eq:13}  implies 
\begin{equation}
  \label{eq:14}
\epsilon_J\leq 1, \quad \epsilon_J \equiv \frac{J}{m^2},
\end{equation}

We can also obtain upper bounds for the total amount of gravitational
radiation emitted by the system. The total energy radiated is given by
$E=m-m_f$ where $m_f$ is the mass of the final Kerr black hole. Since
$m_f^2 \geq J$ we have
\begin{equation}
  \label{eq:15}
  \frac{E}{m}\leq 1- \sqrt{\epsilon_J}.
\end{equation}
In contrast with \eqref{eq:13} and \eqref{eq:14}, the inequality
\eqref{eq:15} involves the complete evolution of the initial data. 

No analytic proof is available so far for \eqref{eq:13}. On the other
hand if the inequality \eqref{eq:13} fails for some initial data then,
these data represent a counter example of weak cosmic censorship.
Beside the trivial example of Kerr initial data, inequality
\eqref{eq:13} has been only studied in the spinning Bowen-York initial
data (cf. \cite{York82}, \cite{Choptuik86b}, \cite{Cook90a}). The
spinning Bowen-York data \cite{Bowen80} are conformally flat and the
second fundamental form is a explicit solution of the momentum
constraint which contain the angular momentum of the data as a free
parameter. The mass of the data has to be computed numerically by
solving the Hamiltonian constraint. Remarkably, these data reach a
limit in $\epsilon_A$ and $\epsilon_J$
\begin{equation}
  \label{eq:16}
0.812 \lesssim\epsilon_A \leq 1 , \quad \epsilon_J \lesssim 0.928.
\end{equation}
$\epsilon_A$ reaches the upper limit only in the nonrotating case,
since in this case the Bowen-York data reduce to the Schwarzschild data.
$\epsilon_J$ cannot reach the limit case $1$ because they are not slices of
Kerr for any choice of the free parameter $J$, even when $J$ goes to
infinity. It appears that the Kerr metric admits no conformally flat
slices (in fact, in \cite{Price00} it has been shown that there does
not exist axisymmetric, conformally flat foliations of the Kerr
spacetime that smoothly reduce, in the Schwarzschild limit, to slices
of constant Schwarzschild time). It is rather easy to construct
conformally flat data with smaller $\epsilon_J$: take the conformal
second fundamental form of Bowen-York $K^{ab}_{BY}$ and add a solution
of the momentum constraint $ K^{ab}_{R}$ such that it has no angular
momentum and such that the square of the sum $K^{ab}_{BY}+ K^{ab}_{R}$
is bigger than the square of $K^{ab}_{BY}$ (these solutions can
be explicitly constructed using Theorem 14 of \cite{Dain99}). Solve
the Hamiltonian constraint with respect to the new second fundamental form
$K^{ab}_{BY}+K^{ab}_{R}$. Then the new data will have bigger
mass and equal angular momentum than the Bowen-York one.

It can be proved that it is not possible to reverse this argument to
produce a data with bigger $\epsilon_J$. Because of this,  one is tempted
to believe that \eqref{eq:16} is the upper limit to all conformally
flat initial data. We will see that this is not the case.

In order to test inequalities \eqref{eq:13}--\eqref{eq:15} in a sharper
way than with the Bowen-York data we need to construct data with
higher $\epsilon_A$ and $\epsilon_J$. To do this, it is natural to consider
perturbations of the Kerr initial data. However there are many ways to
perturb the Kerr initial data, and for each case we have  to deal with
the solvability of a nonlinear elliptic equation in order to get  a
solution of the constraints. We have found a remarkable simple way of
solving this problem. In this article we describe the construction of
a conformally flat initial data in which the second fundamental form
is related to the Kerr second fundamental form in a simple way. The
second fundamental form will be an explicit solution of the momentum
constraint with the following property: It is a conformal rescaling of
the Kerr second fundamental form. These new data can be interpreted as
a conformally flat deformation of the Kerr initial data. 
We find that for the new data
\begin{equation}
  \label{eq:17}
 0.813 \lesssim\epsilon_A\leq1  , \quad \epsilon_J \lesssim 0.932.
\end{equation}

We also find that the total energy radiated is smaller than the
Bowen-York one and satisfies the inequality \eqref{eq:15}.

The initial data presented here is also relevant for the binary
problem. In order to improve the reliability of the results found in
\cite{Baker01b} it is necessary to explore other family of initial data
to know whether these results depend or not on the specific data used;
that is, whether there exist physical properties of the wave form
emitted by a binary system that are invariant under small changes on
the data. It will be possible to measure only this kind of
properties. Recently, other families of initial data has been suggested
(see for instance Ref. \cite{Dain00c}\cite{Dain99b} and
\cite{Bishop98} \cite{Marronetti00} for a Kerr-Schild type). The data
constructed here is as simple as the standard Bowen-York,  they
will be a good candidate for future test and comparison with other
initial data.

In Sec.~\ref{sec:initial} we construct a explicit solutions of the
momentum constraint for conformally flat metrics such that they are
conformal to the Kerr second fundamental form. These solution are
constructed in an invariant way, they will depend on a scalar function
$\omega$ that can be explicitly calculated for the Kerr initial data.
In Sec.~\ref{sec:numerical} we describe the numerical computations of
sequences of initial data and study the $\epsilon_J$ and $\epsilon_A$
for increasing values of $J$.  We then numerically evolve these
initial data to compute the total gravitational energy radiated to
infinity in order to compare it with the Bowen-York solution (and the
vanishing Kerr data). Finally in Sec.~\ref{sec:discussion} we discuss
the extension of our solution to the binary (an multi) black hole
case.

\section{Constraint equations with axial symmetry and the new
  data}\label{sec:initial}  
The standard conformal method for solving the constraints equations for 
maximal initial data is
the following (cf. \cite{Choquet99}, \cite{Choquet80} and the reference
given there). We give a conformal metric $h_{ab}$ and a symmetric
trace free tensor $K^{ab}$ such that
\begin{equation}
  \label{eq:1}
   D_a K^{ab}=0,
\end{equation}
where $D_a$ is the covariant derivative with respect to
$ h_{ab}$. Then,  we solve the following equation for the conformal
factor $\varphi$
\begin{equation}
  \label{eq:2}
  L_{ h} \varphi= -\frac{{K}^{ab}{K}_{ab}}{8\varphi^7},
\end{equation}
where   $L_{ h}= D^a  D_a-  R/8$,  $ R$ is the Ricci scalar of the metric
$ h_{ab}$ and the indexes are moved with  $ h_{ab}$.
 The physical fields  defined by  $\bar{h}_{ab} =
\varphi^4  h_{ab}$ and $\bar{K}^{ab} = \varphi^{-10}{K}^{ab}$ will
satisfy the vacuum constraint equations.  We need to prescribe
appropriate boundary condition to equations (\ref{eq:1}) and
(\ref{eq:2}), we will come back to this point later on.

Remarkable simplifications on (\ref{eq:1}) and (\ref{eq:2}) occur when
$ h_{ab}$ has a Killing vector $\eta^a$. We will assume that $\eta^a$  is
hypersurface orthogonal and we define $\eta$   by $\eta=\eta^a
\eta^b  h_{ab}$.
We analyze first the momentum constraint (\ref{eq:1}) (cf.
\cite{Brandt94a}, \cite{Baker99b}  and  \cite{Dain99b})
Consider the following vector field  $S^a$  
\begin{equation}
  \label{eq:axialve}
  S^a=\frac{1}{\eta} \epsilon^{abc} \eta_b  D_c \omega, \quad
  \pounds_\eta \omega =0, 
\end{equation}
where  $\pounds_\eta$ is the Lie derivative with respect
$\eta^a$ and $\epsilon_{abc}$ is the volume element of $h_{ab}$. 
It follows  that $S^a$ satisfies
\begin{equation}
 \label{eq:J}
\pounds_\eta S^a=0, \quad S^a\eta_a=0, \quad D_a S^a=0.
\end{equation}
Using  the Killing equation $D_{(a}\eta_{b)}=0$,  the fact that
$\eta^a$ is hypersurface orthogonal, (i.e.; it satisfies $D_a
\eta_b=-\eta_{[a}D_{b]} \ln \eta$) and equations (\ref{eq:J}) we
conclude that the  tensor 
\begin{equation}
  \label{eq:axialpsi}
  {K}^{ab}=\frac{2}{\eta} S^{(a} \eta^{b)},
\end{equation}
is trace free and satisfies (\ref{eq:1}).
The square of ${K}^{ab}$ can be written in terms of $\omega$
\begin{equation}
  \label{eq:7}
  {K}^{ab}{K}_{ab}=2\frac{D_c\omega D^c\omega}{\eta^2}.
\end{equation}

Two facts  are important. First, the function $\omega$ is arbitrary. In
particular it does not depend on the metric $h_{ab}$. Second, the
extrinsic curvature of the Kerr 
initial data (in the Boyer-Lindquist coordinates) has the form
(\ref{eq:axialpsi}), and then it has a corresponding function $\omega_K$.  

We can describe now the new data. Let $h$ be the flat metric. Take the
function $\omega_K$. Define ${K}^{ab}$ by (\ref{eq:axialve}) and
(\ref{eq:axialpsi}). Solve (\ref{eq:2}) for the conformal factor with
the appropriate boundary condition. We will obtain a conformally flat
data with a extrinsic curvature that `resemble' the Kerr extrinsic
curvature. In other words, we use the function $\omega$ to construct
out of the Kerr extrinsic curvature a explicit solution of the
momentum constraint (\ref{eq:1}) for flat metric.

In order to write the equations and the boundary conditions explicitly
we introduce spherical coordinates $(r, \theta, \phi)$, and write the
metric in the form \cite{Brill59}
\begin{equation}
  \label{eq:3}
  h = e^{-2q}(dr^2+r^2 d\theta^2)+r^2\sin^2\theta d\phi^2.
\end{equation}
Then, the Hamiltonian constraint (\ref{eq:2}) reads
\begin{equation}
  \label{eq:4}
  \Delta \varphi - \frac{\delta q}{4}\varphi =-\frac{(r\partial_r\omega)^2 +
    (\partial_\theta \omega)^2}{4 r^6 \sin^4\theta \varphi^7},
\end{equation}
where $\Delta$ is the flat Laplacian in the spherical coordinates
$(r, \theta, \phi)$ and $\delta$ is the two-dimensional Laplacian
\begin{equation}
  \label{eq:8}
  \delta q = \frac{1}{r}\partial_r (r\partial_r q) +\frac{1}{r^2}
  \partial^2_\theta q. 
\end{equation}
The Killing vector is given by $\eta^a=(\partial/\partial \phi)^a$,
$\eta=r^2\sin^2\theta$ and $\pounds_\eta \omega = \partial_\phi
\omega=0$.  Note that $q$ and $\omega$ are almost free function. For
$q$ we impose that is regular at the axis. Also, in order to have
solutions for equation (\ref{eq:4}), we need to impose some global
condition on $q$ (cf. \cite{Cantor81}). In our case, both condition
are trivially satisfied since $q$ will be chosen to be zero. For the
function $\omega$ we need that it cancel out the singular denominator
$\sin^4\theta$ in (\ref{eq:4}), in order to obtain solutions which are
smooth in $\theta$. The function $q$ determine the conformal metric,
for $q=0$ we have that the data is conformally flat. The function
$\omega$ determines the extrinsic curvature of the data.

To solve (\ref{eq:4}) we use the following boundary condition
(cf. \cite{Friedrich88,Beig91,Brandt97b} )
\begin{equation}
  \label{eq:9}
  \lim_{r\rightarrow \infty} \varphi =1, \quad  \lim_{r\rightarrow
    0} r\varphi=\frac{m_0}{2}, 
\end{equation}
where $m_0$ is a positive constant called bare mass.
The total angular momentum of the data is given by
\begin{equation}
  \label{eq:130}
  J= -\frac{1}{8\pi} \int_S S_a  n^a \, dS,
\end{equation}
where $S$ is any closed two-surface which enclose the origin. This
equation can be written in a remarkable simple form in terms of
$\omega$
\begin{equation}
  \label{eq:132}
  J=\frac{1}{4}(\omega(r,\theta=\pi) -\omega(r, \theta=0)).
\end{equation}

Let us mention some examples. 
The spinning Bowen-York\cite{Bowen80} initial data is obtained as a
solution of equation (\ref{eq:4}) with  $q=0$, $m_0$ an arbitrary
constant, and $\omega$ given by
\begin{equation}
  \label{eq:141}
  \omega_{BY}=J(\cos^3\theta-3\cos\theta).
\end{equation}
The  Kerr initial data is obtained as a solution  of
(\ref{eq:4}), where  the function $q$ is given by 
\begin{equation}
  \label{qKerr}
  e^{-2q_{K}} =  \frac{\Sigma}{r_{BL}^2+a^2+
\frac{2m_Ka^2 r_{BL}\sin^2\theta}{\Sigma}}, 
\end{equation}
where
\begin{equation}
  \label{eq:143}
\Sigma=r^2_{BL} +a^2 \cos^2\theta,  
\end{equation}
\begin{equation}
  \label{eq:144}
  r_{BL} =r +m_K+\frac{m_K^2-a^2}{4 r},
\end{equation}
and  $m_K$  and $a$ are the Kerr parameters, and $r_{BL}$ is the usual
Brill-Lindquist radial coordinate. 
The function $\omega$ is given by
\begin{equation}
  \label{eq:142}
  \omega_K=\omega_{BY}- \frac{m_Ka^3\sin^4\theta \cos\theta}{\Sigma},
\end{equation}
where, $J=m_Ka$. And the parameter $m_0$ in Eq.~(\ref{eq:9}) is given by
\begin{equation}
  \label{eq:10}
  m_0=\sqrt{m_K^2-a^2}.
\end{equation}
We see that the second fundamental form  we are considering can be interpreted
as a correction ${\cal O}(J^3)$ (and higher) to that of Bowen and York as is
directly reflected in the form of  $\omega_K$.

The new data is $q=0$, and $\omega$ and $m_0$ given by Eqs.\ 
(\ref{eq:142}) and (\ref{eq:10}) respectively. The explicit components of
the conformal second fundamental form are given by 
\begin{align} \label{eq:KerrKij}
K_{r\phi}&= \frac{am_K
\left[
(r_{BL}^2-a^2)\Sigma+2r_{BL}^2(r_{BL}^2+a^2)
\right]
}{
r^2\Sigma^2 
}\sin^2\theta,\\
K_{\theta\phi}&= \frac{
-2a^3m_Kr_{BL}}
{\Sigma^2
}\left(1-\frac{m_K^2-a^2}{4r^2} \right)\cos\theta \sin^3\theta.\nonumber
\end{align}

Note that $m_0$ is chosen such that we recover the Kerr initial data
if we insert in equation (\ref{eq:4}) the corresponding function
$q_K$.

Since $q=0$, in this
case the boundary condition (\ref{eq:9}) can be achieved by the ansatz 
$\phi=1+m_0/(2r)+u$, where the function $u$ is bounded at the origin
and satisfies the regular equation in $\mathbb{R}^3$
\begin{equation}
  \label{eq:11}
 \Delta u =-\frac{r((r\partial_r\omega_K)^2 +
    (\partial_\theta \omega_K)^2)}{4  \sin^4\theta (r+m_0/2+u)^7},  
\end{equation}
with the boundary condition $\lim_{r\rightarrow \infty} u =0$.
This allows to use the `puncture' treatment in numerical
implementations of these initial data~\cite{Brandt97b}.
Existence and uniqueness of positive solution of the elliptic equation
\eqref{eq:11} can be proved by standard methods (see for example
\cite{Choquet99} and \cite{Dain99} and reference therein). 

Finally, we want to prove that these data have an isometry
defined by an inversion map through a sphere of radius
$m_0/2$. Inversion transformation in the context of black holes
initial data have been studied in \cite{Bowen80}. The inversion map
$I$ is defined in spherical coordinates by
\begin{equation}
  \label{eq:18}
  r\rightarrow \left(\frac{m_0}{2}\right)^2 \frac{1}{r}, \quad \theta
  \rightarrow \theta, \quad \phi \rightarrow  \phi.
\end{equation}
We want to prove the following assertion: if the functions $\omega$
and $q$ satisfy
\begin{equation}
  \label{eq:19}
  I\circ q= q, \quad I\circ \omega= \omega,
\end{equation}
(that is, they are invariant under $I$); then the physical fields
satisfy
\begin{equation}
  \label{eq:20}
  (I^*\bar h)_{ab}=\bar h_{ab}, \quad (I^*\bar K)_{ab}=-\bar K_{ab}.
\end{equation}
In the examples presented here, one easily check that $\omega$
and $q$ satisfy Eq. (\ref{eq:19}), since their radial dependence is
given in terms of $r_{BL}$ defined by Eq. (\ref{eq:144}), which
satisfy $I\circ r_{BL}=r_{BL}$. 

In order to prove the assertion, we first calculate the transformation 
of the conformal fields under $I$, from Eqs. (\ref{eq:axialpsi}) and
(\ref{eq:3}) we obtain
\begin{equation}
  \label{eq:21}
 (I^* h)_{ab}=\left(\frac{m_0}{2r}\right)^4h_{ab}, \quad (I^*
 K)_{ab}=-\left(\frac{m_0}{2r}\right)^{-2} K_{ab}. 
\end{equation}
Then, using Eq. (\ref{eq:2}) we obtain
\begin{equation}
  \label{eq:22}
  I\circ \varphi=\left(\frac{m_0}{2r}\right)^{-1}\varphi. 
\end{equation}
From Eqs. (\ref{eq:21}) and (\ref{eq:22}) follow Eqs. (\ref{eq:20}). 
This property of the data is used in the numerical calculations, as we 
will see in the following section.

\section{Numerical Results}
 \label{sec:numerical}

The 2D fully nonlinear evolutions have been performed with a code,
{\sl Magor}, designed to evolve axisymmetric, rotating, highly
distorted black holes, as described in
Ref.~\cite{Brandt94a,Brandt94b}.  This nonlinear code solves the
complete set of Einstein equations, in axisymmetry, with maximal
slicing, for a rotating black hole.  The code is written in a
spherical-polar coordinate system, with a rescaled logarithmic radial
coordinate that vanishes on the black hole throat.  An isometry
operator is used to provide boundary conditions on the throat of the
black hole. The lapse is chosen to be antisymmetric across the throat.
 The shift vector components are employed to
keep off diagonal components of the metric zero, except for the
$g_{\theta\varphi}$.  The initial data described in
Sec.~\ref{sec:initial} above are provided through a fully nonlinear,
numerical solution to the Hamiltonian constraint.

\begin{figure}
\includegraphics[width=3.0in]{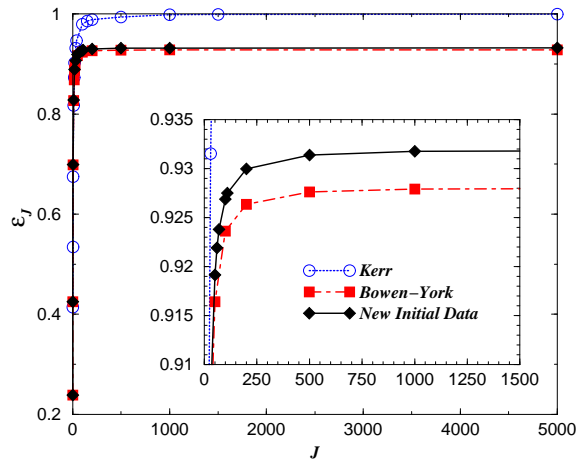}
\caption{\label{fig:adm} Rotation parameter $\epsilon_J$ as
  a function of the angular momentum $J$  for the new data, Kerr,
  and Bowen-York holes along the curve, keeping fixed the free
  parameter $m_0=2$. The curve for Kerr 
can be obtained analytically by equation \eqref{eq:10}.}
\end{figure}

In Fig.\ \ref{fig:adm} we plot the curves of constant bare mass
$m_0=2$ for initial data corresponding to rotating holes of Kerr,
Bowen-York and the new ones introduced in this paper.  Interestingly
enough it was shown in Refs.\ \cite{Choptuik86b,Cook90a} that the
Bowen-York hole reach a maximum rotation parameter of
$\epsilon_J\approx 0.928$ when $J$ goes to infinity. 
 We have reproduced this curve as a function
of $J$  reaching values of
$J\approx10,000$. For Kerr holes, along the curve $m_{0}=constant$ we
have equation \eqref{eq:10}; the extreme Kerr $a/m_K=1$   correspond  
to the limit $J\rightarrow \infty$.   We find for
the data presented in this paper that the maximum lies near
$\epsilon_J=0.932$, which is higher than the Bowen-York maximally
rotating hole. This allows to study black hole evolutions from values
closer to the maximally rotating Kerr ones.

\begin{figure}
\includegraphics[width=3.0in]{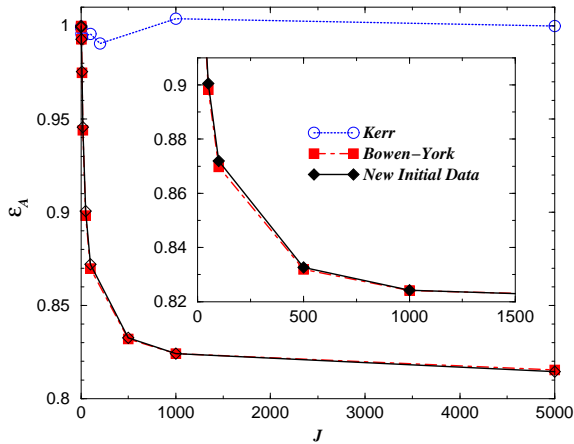}
\caption{\label{fig:a} The horizon area parameter $\epsilon_A$ as
  a function of the angular momentum $J$ for the new data, Kerr, and
  Bowen-York holes along the curve, keeping fixed the free parameter
  $m_0=2$. For the Kerr initial data the exact value is
  $\epsilon_A=1$.}
\end{figure}

\begin{figure}
\includegraphics[width=3.0in]{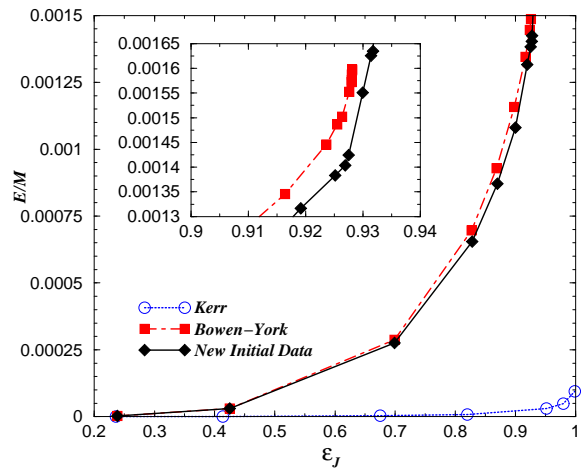}
\caption{\label{fig:rad} Radiation content of a single rotating black
hole given by the new, Kerr and Bowen-York initial data. The upper
bound  given by \eqref{eq:15} is in this case $ \approx  0.031$. }
\end{figure}

The code is able to evolve such data sets for time scales of roughly
$t\leq 100m$, and study physics such as location and evolution of
apparent horizons and gravitational wave emission.  We have used
typical grid sizes of 300 radial by 39 angular zones and extracted
waves at the radial location $r_{obs}=15m$.  The results are displayed
in Fig. \ref{fig:rad} and clearly show that the new data has less
radiation content than the spinning Bowen-York holes.  It produces
roughly ten percent less the total radiated energy, $E$.  Also this
plot shows that inequality \eqref{eq:15} is satisfied, the upper
bound given by \eqref{eq:15} is in this case $ \approx  0.031$
while the maximum of the total energy radiated is $\approx 0.0015$.

For completeness we have
plotted the evolution of Kerr initial data, for which the outgoing
radiation should strictly vanish, as measure of the numerical error
of our evolutions; and for comparison with future work we present the
numerical parameters of our initial data family in Table~\ref{JmAE}.

\begin{table}
\caption{Initial spin $J$, ADM mass $m$, apparent horizon mass $m_{AH}=\sqrt{A/16\pi}$,
and energy radiated of the new family of initial data}
\begin{tabular}{llll}\label{JmAE}

$J/(m_0/2)^2$   &        $m/(m_0/2)$     &       $m_{AH}/(m_0/2)$        &       $E/m$
\\ \colrule
1       &2.0478         &2.047          &2.75e-6         \\
2       &2.1694         &2.169          &2.99e-5         \\
5       &2.674          &2.668          &0.00028         \\
10      &3.475          &3.452          &0.00066         \\
20      &4.743          &4.686          &0.00100         \\
50      &7.375          &7.242          &0.00131         \\
100     &10.386         &10.165         &0.00140         \\
500     &23.169         &22.580         &0.00162         \\
1000    &32.760         &31.899         &0.00163         \\
5000    &73.245         &71.248         &0.00191         \\
10000   &103.583        &100.74         &0.00187         \\
\colrule
\end{tabular}
\end{table}

\section{Discussion}
 \label{sec:discussion}
On the light of the two aspects studied in this paper we observe
that the new data improves on the Bowen-York one leading to less
spurious radiation and allowing a representation of higher rotating
black holes while keeping the simplicity of the solutions, namely
the explicit analytic form of the extrinsic curvature and conformal
flatness of the three-geometry. This proves that even within the
conformally flat ansatz one can look for astrophysically more realistic
initial data.
Also the new data satisfies inequalities \eqref{eq:13}, \eqref{eq:14}
 and \eqref{eq:15}, supporting the validity of weak cosmic censorship.

\begin{figure}
\includegraphics[width=2.5in]{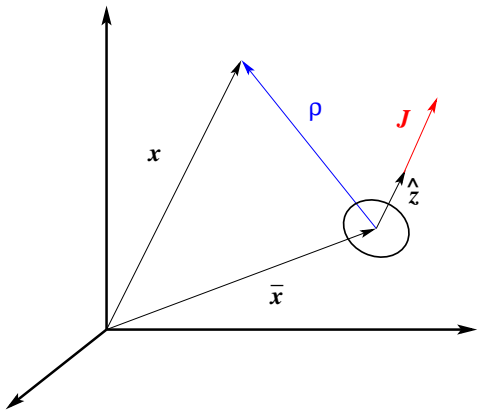}
\caption{\label{fig:nBH}
Geometry in the conformal (Cartesian) space of the
location and orientation of a generic nth spinning black hole.}
\end{figure}

We want to discuss now the generalization of these data for multiple
black holes. In general, if the conformal metric $h_{ab}$ admits only
one axial Killing vector $\eta^a$ the only freedom left is the choice
of the origin in the $z=\cos\theta/r$ coordinate. By superposing different
solutions of the momentum constraint of the form \eqref{eq:axialpsi}
such that they are singular at different points in the axis we will
obtain a multiple black hole solution for a general axially symmetric
metric. The spin of all of the black holes will point in the direction
of the axis.  However, when $h_{ab}$ is chosen to be the flat metric,
then we have that a rotation about any axis is a Killing vector. Hence
it is straightforward to generalize the data presented here to include
multiple black holes in arbitrary location and with spin pointing in
arbitrary direction. For completeness we will write explicitly the
expression for the general solution of the momentum constraint, for
flat metrics, with arbitrary origin and with spin pointing in
arbitrary direction. The general Killing vector for the flat metric
can be written as
\begin{equation}
  \label{eq:5}
  \eta^a=\epsilon^{abc}\hat z_b (x_c -\bar x_c), \quad \eta=\rho^2-
  (\hat z_c(x^c-\bar x^c))^2,
\end{equation}
where $x_c$ are Cartesian coordinates, $\bar x^c$ and $\hat z^a$ are  arbitrary
constant vectors (we chose $\hat z^a$ to be a unit vector) and
$\rho^2=(x^c-\bar x^c)(x_c-\bar x_c)$. The vector $\bar x^c$ represent
the new origin and $\hat z^a$ the new axis. The new coordinate $\theta'$
with respect to the axis $\hat z^a$ is given by
\begin{equation}
  \label{eq:6}
  \cos \theta'= \hat z_c(x^c-\bar x^c).
\end{equation}
Fig.~\ref{fig:nBH} displays explicitly those vectors.

Then, the desired expression for $K^{ab}$ is given again by
Eqs. \eqref{eq:axialpsi}, \eqref{eq:axialve} and  \eqref{eq:142} but
we use  in  equations \eqref{eq:axialpsi} and  \eqref{eq:axialve} the
expression for  
$\eta^a$ and  $\eta$ given by \eqref{eq:5},
and we replace in \eqref{eq:142} $\theta$ by $\theta'$ given
by \eqref{eq:6}.

\begin{acknowledgments}
We wish to thank M. Campanelli and J. Pullin for motivating
discussions and to M. Mars and W. Simon for
illuminating discussions regarding  inequality \eqref{eq:13}. 
\end{acknowledgments}


\end{document}